\documentclass[unnumsec,webpdf,contemporary, large, numbered]{oup-authoring-template}

\usepackage{subfig}

\usepackage{booktabs}
\usepackage[most]{tcolorbox}
\usepackage{comment}
\tcbuselibrary{breakable,skins}
\usepackage{fvextra}
\graphicspath{{Fig/}}

\usepackage{soul}
\usepackage{color}

\theoremstyle{thmstyleone}%
\theoremstyle{thmstyletwo}%
\theoremstyle{thmstylethree}%

\begin{document}

\journaltitle{Preprint}
\DOI{DOI added during production}
\copyrightyear{YEAR}
\pubyear{YEAR}
\vol{XX}
\issue{x}
\access{Published: Date added during production}
\appnotes{Paper}

\firstpage{1}


\title[The Epi-LLM Framework]{The Epi-LLM Framework: probing LLM behavioral priors through epidemiological agent-based models}

\author[1,$\dagger$]{Petra Ferencz\ORCID{0009-0001-4881-7381}}
\author[1,2,$\ast$,$\dagger$]{Ava Keeling\ORCID{0009-0000-9665-7864}}
\author[1,$\dagger$]{Tobias O'Keefe\ORCID{0009-0007-4738-1065}}
\author[1,$\dagger$]{Lorenzo Stigliano\ORCID{0009-0001-9344-8212}}
\author[3]{Francesco Di Lauro}
\author[4,5, $\ddagger$]{Andres Colubri}
\author[3,6, $\ddagger$]{Jasmina Panovska-Griffiths\ORCID{0000-0002-7720-1121}}

\address[1]{\orgdiv{Big Data Institute, Li Ka Shing Center for Health Information and Discovery}, \orgname{University of Oxford}, \orgaddress{Oxford, \country{United Kingdom}}}

\address[2]{\orgdiv{Leverhulme Centre for Demographic Science, Nuffield Department of Population Health}, \orgname{University of Oxford}, \orgaddress{Oxford, \country{United Kingdom}}}

\address[3]{\orgdiv{Pandemic Sciences Institute, Nuffield Department of Medicine}, \orgname{University of Oxford}, \orgaddress{Oxford, \country{United Kingdom}}}

\address[4]{\orgdiv{Department of Genomics and Computational Biology}, \orgname{UMass Chan Medical School}, \orgaddress{\country{USA}}}

\address[5]{\orgdiv{Broad Institute of Harvard and MIT}, \orgaddress{\country{USA}}}

\address[6]{\orgdiv{The Queen's College}, \orgname{University of Oxford}, \orgaddress{Oxford, \country{United Kingdom}}}

\corresp[$\ast$]{Corresponding author. \href{mailto:ava.keeling@stx.ox.ac.uk}{ava.keeling@stx.ox.ac.uk}\\}

\corresp[$\dagger$]{These authors all contributed equally.\\}

\corresp[$\ddagger$]{Senior authors contributed equally.}

\received{Date}{0}{Year}
\revised{Date}{0}{Year}
\accepted{Date}{0}{Year}


\abstract{Human behaviour during epidemics affects infectious disease dynamics, but quantifying this remains deeply challenging. Here we introduce the Epi-LLM framework: a novel integration of agent-based modelling, real-life epigames, and large language models (LLMs) in which a synthetic society of agents reasons and adapts dynamically over an outbreak contact network. Comparing synthetic agent behaviour against a no-intervention SEIR baseline and human participant data from the AUIB epigame study, we find that LLM agents across four different architectures reduced peak active infections, with quarantine compliance peaking at 58–65\% on day six of the 15-day simulation. A binomial generalised linear model showed that perceived health severity was the strongest predictor of quarantine behaviour ($\beta = 0.33, p = 0.002$), yielding a pseudo-R² of 0.055, comparable to the 0.072 observed in the human trial. LLM architecture is a key determinant of epidemic dynamics: low-variance architectures offer greater internal validity for testing behavioural rules, while high-variance models may better represent real-world decision-making. Geographic labels alone do not induce culturally differentiated behaviour; explicit attitudinal parameterisation is required. This proof-of-principle work lays the groundwork for deploying the Epi-LLM framework as a scalable, risk-free simulation environment for pandemic preparedness research.
} 


\keywords[Abbreviations]{American University of Iraq-Baghdad (AUIB), Large Language Model (LLM), susceptible-infected-recovered (SIR), susceptible-exposed-infected-recovered (SEIR), Generative Agent-Based Model (GABM), Agent-Based Model (ABM)}

\boxedtext{Key Messages}{
\begin{itemize}
\item LLM architecture is a key determinant of simulated epidemic dynamics: lower-variance architectures provide greater internal validity for testing specific behavioural rules, while higher-variance models better capture the stochastic nature of real-world decision-making.
\item LLM agents reproduce the broad threat-driver compliance patterns observed in human epigame participants, with perceived health severity as the strongest predictor of quarantine behaviour; modifying incentive structures substantially alters epidemic trajectories.
\item Geographic labels alone do not induce culturally diferentiated behaviour in LLM agents; meaningful cultural variation in generative agent-based models requires explicit parameterisation of attitudinal and normative factors beyond simple location identifiers.
\end{itemize}}

\maketitle

\section{Introduction}

Human behaviour is a critical determinant of infectious disease dynamics. Yet quantifying the impact of dynamical behavioural changes, voluntary quarantining, vaccination uptake, and social distancing, remains deeply challenging in epidemiological modelling, both due to observational constraints and the complexity of predicting human decision-making. Traditional epidemiological methods, such as the susceptible-infected-recovered (SIR) model \citep{kermack1927contribution}, rely on simplifying assumptions that fail to capture the complex, adaptive interplay between public behaviour and pathogen dynamics. Agent-based models (ABMs) \citep{tracy2018agent} have addressed this by simulating individual-level behaviour; large-scale ABMs have been particularly influential in pandemic planning including projecting the impact of non-pharmaceutical interventions during COVID-19 \citep{ferguson2020report} and influenza pandemic modelling \citep{ferguson2006strategies}.

As a result, there is growing recognition that machine learning can address limitations of traditional epidemiological modelling \citep{ye2025integrating}, with applications ranging from model calibration to intervention assessment. For instance, deep learning has been used to identify optimal vaccination strategies within agent-based simulations \citep{jian2017applying}, a direction adjacent to, but distinct from, the behavioural simulation approach explored here. Large language models (LLMs) have recently been incorporated into pandemic science in several distinct ways. Beyond aiding modelling approaches, several works focus on their use as decision-support tools, examining how they can accelerate reaction to emerging outbreaks \citep{kaur2025ai}, evaluating their application in infection prevention and control within hospital settings to assist clinicians \citep{wong2025comparative}, and surveying the broader landscape of such tools designed for outbreak response \citep{rizzo2024future}. Other works explore the use of LLM agents as autonomous research systems. For example, \citep{samaei2026epidemiqs} introduces an end-to-end agent framework for epidemic modelling and analysis, designed to orchestrate the full modelling pipeline in the manner of an AI researcher, analogous to recent developments in automated software engineering and scientific discovery \citep{yang2024swe, lu2024ai, boiko2023emergent}.

LLM agents have also been deployed in healthcare simulation. For example, Agent Hospital \citep{li2024agent} uses fully LLM-driven agents in a hospital setting, although that work focuses on clinical training rather than population-level epidemic dynamics.

Despite this progress, the use of language models as the behavioural engine within agent-based simulations, where individual agents reason, communicate, and adapt dynamically over the course of an outbreak, remains comparatively underexplored. This intersection, broadly referred to as generative agent-based models (GABMs), has shown significant potential in modelling real-life behaviours, offering a novel paradigm for understanding complex social systems \citep{lu2024llms, park2023generative}. GABMs have demonstrated strong empirical grounding in adjacent domains. For instance, they have been used to simulate opinion dynamics over networks \citep{chuang2024simulating}, and pairing agents with qualitative interview transcripts has been shown to replicate human survey responses with high fidelity while reducing demographic biases \citep{park2024generative}. These results suggest that anchoring agent behaviour in real-world data can yield simulations that closely reflect the diversity of human decision-making. More directly relevant to epidemic modelling, \citep{choi2025infected} used GABMs to explore behavioural differences between agents exposed to outbreak news and those who were not, finding that agents reading about the epidemic significantly reduced their movement and social engagement, showing that GABMs can capture nuanced, context-sensitive behavioural responses. Similarly, \citep{williams2023epidemic} demonstrates that LLMs can generate plausible epidemic behaviours within an agent-based setting. 

However, these works are limited by their reliance on a single model architecture, small agent populations, and the absence of empirical grounding in real-world behavioural data. Moreover, ABMs such as Starsim \citep{starsim2026} and Covasim \citep{kerr2021covasim} have proven valuable for simulating epidemic dynamics during COVID-19, capturing individual-level heterogeneity in contact patterns and intervention responses. However, these models have largely been constrained by the absence of empirical behavioural data relying instead on inferred parameters to represent how individuals make health-related decisions. Epigames, i.e. smartphone games that simulated epidemics in real-world social environments \citep{colubri2026understanding}, can generate behavioural data from decisions during the game as well as attitudinal data from surveys before and during the epigame. Analysis of the data from a recent epigame ran at the American University of Iraq–Baghdad (AUIB) demonstrated that participants real-life health beliefs closely correlate with their in-game quarantine preferences \citep{colubri2026app}. Here, we extend this study to further address the gap of empirical behavioural data by parameterising LLM agents using real-world survey responses to construct a ``synthetic society'' in which agents act and interact over a contact network, with outbreak dynamics emerging as a function of human preferences and decision-making patterns. To this end, we develop the Epi-LLM framework: a novel integration of agent-based modelling, real-life epigames, and LLMs in which agents reason and adapt dynamically over the course of an outbreak. Through a number of scenarios, we explore the effect of LLM architecture and agent behaviour on epidemic dynamics and trajectory and examine how similar the behaviour of the real-life and LLM agents is.


\section{Methods}

\subsection{Epi-LLMs simulation Framework}

The simulation framework was built on top of \textit{Starsim}, an open-source platform for agent-based modelling of infectious disease dynamics \citep{starsim2026}. Starsim expands the earlier design of \textit{Covasim}, an agent-based Covid-19 modelling tool that was widely and successfully used throughout the pandemic to simulate transmission dynamics and evaluate the effects of non-pharmaceutical interventions such as testing, isolation and vaccination \citep{kerr2021covasim}. Unlike commonly used population compartmentalised models, such as classical SEIR models, agent-based models explicitly represent heterogenous individuals and their interactions over dynamic contact networks. This enables the modelling of behavioural adaptation, local contact structure, and stochastic decision making, all of which are central to this work. Starsim's modular support for dynamic contact networks, disease progression, and intervention pipelines makes it well suitef for LLM-driven behavioural agents. To support reproducibility, all code and implementation details are publicly available at \url{https://github.com/pferencz23/starsim}.

\begin{figure*}[h]
    \centering
    \includegraphics[width=\linewidth]{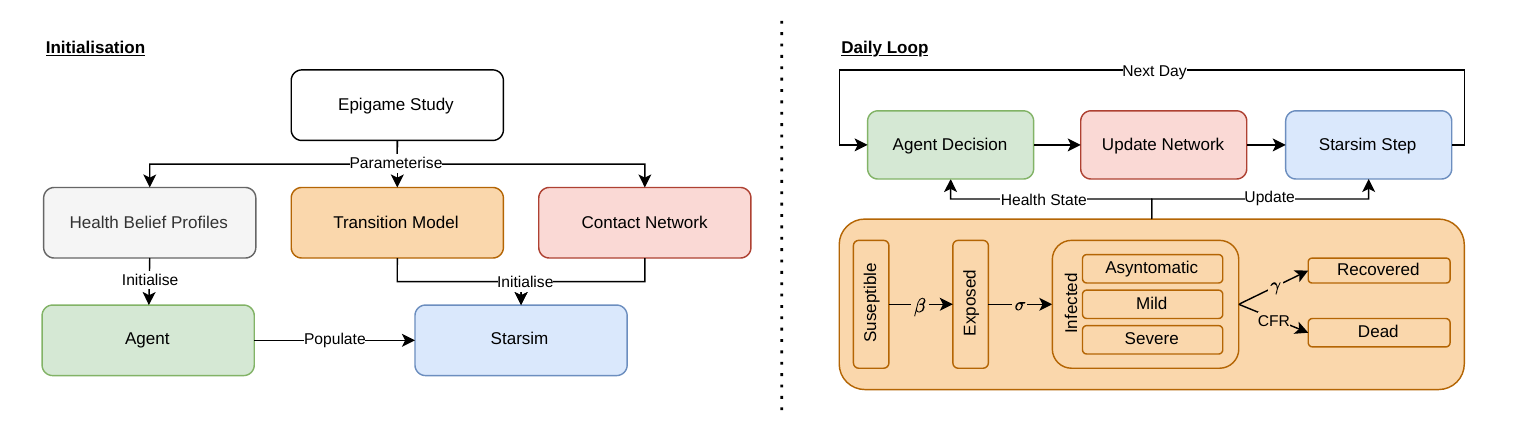}
    \caption{Overview of the Epi-LLM framework. Initialisation phase: agents are parameterised using real-world Health Belief Model survey responses from the AUIB epigame study, a contact network is generated from a fitted log-normal degree distribution, and initial disease prevalence. Daily simulation loop: each agent observes its current health status, local symptomatic prevalence, and point balance before submitting a quarantine decision via LLM prompt; the contact network is rewired, and the Starsim SEIR update propagates transmission and disease progression across the network. CFR = Case Fatality Rate. Together, the panel shows how empirical behavioural data and LLM reasoning are embedded within a standard compartmental disease model.}
    \label{fig:framework}
\end{figure*}

An overview of the framework is provided in Figure~\ref{fig:framework}, which illustrates the initialisation phase and the daily simulation loop in detail.

To ground the simulation in realistic behavioural dynamics, we used the AUIB study as an illustrative epigame example, providing the social context for outbreak-related decisions. This allowed LLM agents to operate in a socially grounded environment, while pathogen dynamics were modelled separately in Starsim.

\subsubsection{Starsim Transmission Model and Epidemiological Parameters}

The pathogen was modelled using a modified version of Starsim’s built-in SIR module \citep{weiss2013sir}, extended to include an additional \emph{exposed} compartment, thereby yielding an SEIR compartmental model \citep{biswas2014seir}. To reflect the features of the AUIB study, the model also allowed infected agents to experience different levels of disease severity: asymptomatic, mild, or severe. Disease severity was assumed to influence mortality risk. Pathogen transmission was simulated over 15 days, matching the duration of the specific epigame. The epidemiological parameters used in the simulation are summarised in Table~\ref{tab:epidemiological_params}. 
 
\begin{table}[ht]
    \begin{center}
    \caption{Disease severity and mortality parameters used in the simulation.}
    \begin{tabular}{ll}
        \toprule
        Parameter & Value\\
        \midrule
        Initial disease prevalence  & 1\% \\
        Infectiousness ($\beta$) & 0.0907 per hour \\
        Rate of progression from exposed to infected ($\sigma$) & 0.1 per hour \\
        Recovery rate ($\gamma$) & 1/77 per hour \\
        \midrule
        Death rate - asymptomatic (30\%) & 0.00 \\
        Death rate - mild disease (42\%) & 0.25 \\
        Death rate - severe disease (28\%) & 0.70 \\
        \botrule
    \end{tabular}
    \label{tab:epidemiological_params}
    \end{center}
\end{table}

These time scales are reconciled as follows. The 15-day simulation matches the duration of AUIB epigame, with LLM agents making one binary quarantine decision per simulated day. Disease transmission is computed at an hourly resolution within Starsim, with the infectiousness parameter $\beta = 0.0907$ per hour calibrated to produce realistic outbreak dynamics over this window. Contact durations of 10 secods (the median of the empirical distribution) reflect the transient Bluetooth-sensed contacts recorded during the AUIB study and are used to set the edge-persistence parameter in the dynamic network. The simulation corresponds to the 2024 AUIB cohort data.

\subsubsection{Agents interactions and contact network}

\begin{figure*}
\centering

\subfloat[Empirical degree distribution from the AUIB study.\label{fig:degree_real}]{
    \includegraphics[width=0.45\linewidth]{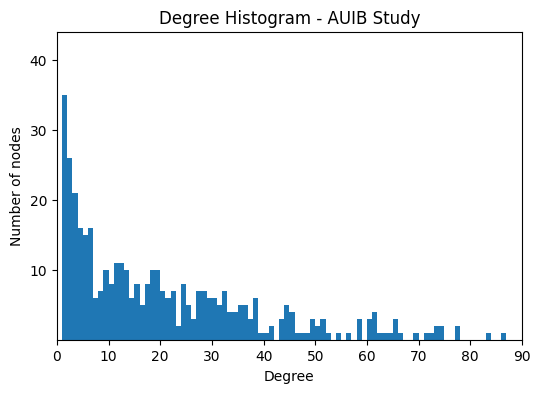}
}
\hfill
\subfloat[Degree distribution from the fitted log-normal model.\label{fig:degree_sim}]{
    \includegraphics[width=0.45\linewidth]{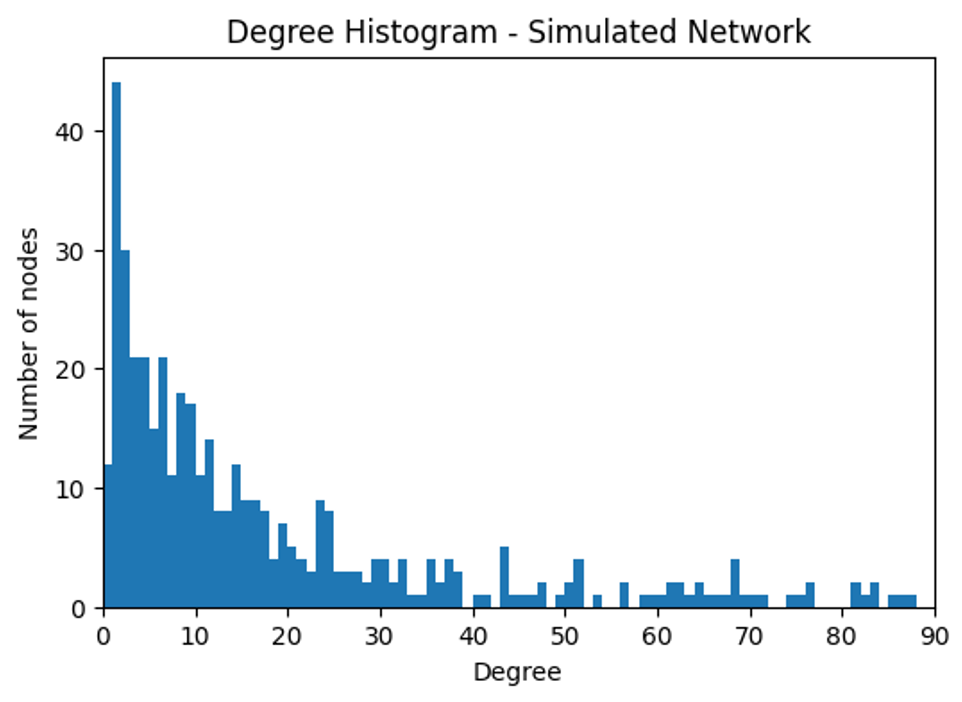}
}

\caption{Comparison of empirical and simulated degree distributions. Panel A) shows the observed degree distribution from the AUIB study, while panel B) shows the distribution obtained from samples of the fitted log-normal model. The similarity in skewness and spread indicates that the log-normal distribution provides a good approximation to the empirical contact structure, justifying its use as the basis for agent contact networks in all simulations.}
\label{fig:degree_comparison}
\end{figure*}

The RandomNet module within the Starsim framework was used to generate a realistic contact network between agents. This module constructs random graphs with a user-specified mean degree and edge duration, after which all connections are rewired. We chose a dynamic rather than a static network to closer reflect how social networks naturally occur, with connections happening and finishing over the course of the experiment. If we were to use a static network, where edges do not change at all over the course of the epigame, we would have seen rapid saturation of infection across fixed connections.

We modelled contacts using a random network as a neutral baseline capturing average connectivity and short-term mixing. While this approach does not capture higher-order structure such as long-term friendships and community clustering, such features are less relevant in this context due to the short-lived proximity based interactions recorded via Bluetooth sensing. Over these short time windows, contact patterns can approximate random mixing \citep{holme2012temporal}. 

We parameterise the network using information from epigames, and illustrate this by having a network structure from the AUIB study. This means that the observed degree distribution (Figure~\ref{fig:degree_real}) exhibits a strong right skew, with most individuals having few contacts and a small number acting as high-degree outliers. To capture this, we modelled node degrees using a log-normal distribution, a common approach in social contact networks \citep{danon2012social}. The distribution was parameterised with a mean and standard deviation of the log-normal distribution itself (in units of contacts per agent, not the log-scale parameters of the underlying normal), chosen so that the simulated degree distribution closely matches the empirical AUIB data. Samples from the fitted model closely reproduce the empirical distribution (Figure~\ref{fig:degree_sim}), indicating good agreement in both skewness and dispersion.

We also examined the distribution of contact durations, which was highly skewed. To avoid sensitivity to extreme values, we used the median contact duration of 10 seconds as the edge persistence parameter in the dynamic network.

\subsubsection{Decision Framework}

Once per day, agents were asked to make a binary decision: enter quarantine for that round or remain active in the network using the prompt described in Appendix~\ref{app:Prompt Engineering Strucutre}. This was implemented using a custom intervention module within the Starsim framework. Mirroring the design of the AUIB study, this framework imposed a trade-off between short-term reward and long-term risk and included group randomisation with a varying reward incentive. Agents who chose to quarantine received a reduced daily reward (5 points) but were guaranteed protection from infection for that round. Agents who remained active were exposed to potential transmission from nearby contacts, but could earn higher rewards, 10 points for Group A participants, and 15 points for Group B. This came at the cost of the infection risk and its associated penalties. If an agent dies they lose all accumulated points and are removed from the network.

\subsection{LLM Agent Architectures}

All LLMs were accessed via the \textit{OpenRouter} API \citep{openrouter2026}, which provides a unified interface enabling consistent deployment and comparison across model architectures.

\subsubsection{Model Selection}

To investigate the effect of model architecture on the simulated epidemic behaviour, we instantiated agents using four LLMs: DeepSeek V3 \citep{liu2024deepseek}, Llama 3 70B \citep{grattafiori2024llama}, Nemotron 120B (Nemotron-3-Super) \citep{blakeman2025nvidia}, and GPT-OSS 120B \citep{agarwal2025gpt}. These models were selected to contrast two architectural families, dense transformer and Mixture-of-Experts (MoE), across a range of parameter scales (70B to 120B), allowing assessment of whether architectural family rather than scale drives variation in quarantine behaviour and epidemic trajectories. Each model received an identical prompt structure designed to reproduce the behavioural and informational conditions of the AUIB study \citep{colubri2026understanding}. The prompt supplied agents with epidemiological, behavioural, and network-level context, ensuring that observed differences in quarantine behaviour and epidemic outcomes were attributable to model architecture rather than prompt asymmetry.

Agent behavioural profiles were parametrised using Health Belief Model belief scores derived from AUIB participant survey data, capturing perceived susceptibility, severity, benefits, and barriers. Full score construction and preprocessing details are provided in Appendix~\ref{app:behavioural_params}.

\subsubsection{Geographic Framing}
The agents were assigned to one of four geographic locations: China, Iraq, Kenya, or the United Kingdom. These were selected to span four world regions with markedly different reported compliance norms and public-health governance contexts, enabling a first-pass test of whether simple geographic labels elicit culturally differentiated responses from LLM agents. Full detauls of the augmented prompt are in Appendix~\ref{app:prompt_template_geography}.

\subsection{Experimental Design and Evaluation}

\subsubsection{Statistical analysis}
To evaluate how the health belief inputs influenced the LLMs’ decision to quarantine and to compare this with the behaviour of human participants in the AUIB study, two generalised linear models were fitted to the simulation data.

A binomial GLM was used to analyse the proportion of days each agent spent in quarantine. This models the probability that an agent was quarantining on any given day during its survival period. The model included group assignment (high- vs. low-quarantine cost) as a categorical predictor and the four perceived belief variables—perceived health severity, perceived infection risk, quarantine self-efficacy, and response efficacy—each centred at zero, as continuous predictors.

A Poisson regression model was then fitted with each agent’s quarantine count as the outcome variable, offset by the number of days the agent survived in the simulation. Group membership was included as an interaction term with each of the four questionnaire-derived predictors. This allowed us to test whether the associations between health-belief scores and quarantine behaviour differed between the low-cost group (Group A) and the high-cost group (Group B).


\subsubsection{Alternative Incentive Structure}

\label{section:diff_cost}

We explored alternative incentive structures by implementing a new scoring system within the game. This consisted of each agent earning a baseline of 5 points each day plus an additional 3 points for every contact made. In this simulation, the ``cost'' of quarantining is forgoing out on the extra points the agent may have gotten through contacts. This method was intended to simulate the benefits that people may get from interacting with people in their day-to-day life, which may ultimately be a reason for someone choosing to not quarantine when they should have. 

\subsubsection{Decision Consistency Analysis}
\label{section:consistency}

We also explored if agents behave inconsistently under identical conditions. Observed differences in epidemic trajectories across architectures could reflect stochastic action noise rather than meaningful differences in behaviours (details in Appendix~\ref{app:decision_consistency}). 


\section{Results}

\subsection{Effect of LLM Architecture on Epidemic Dynamics}
\label{llmarchitectyre_section}

\begin{figure*}
\centering

\subfloat[Active infections over time. \label{fig:a_active_infections_model}]{
    \includegraphics[width=0.3\linewidth]{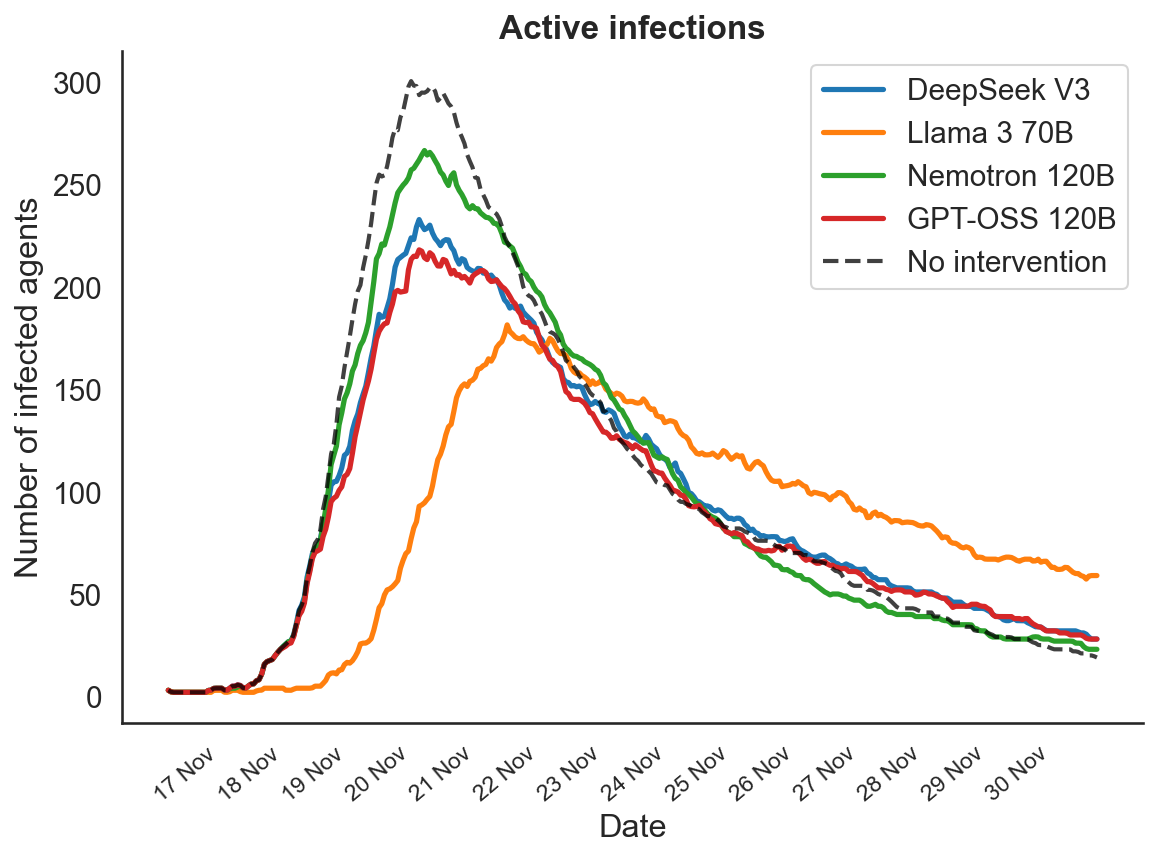}
}
\hfill
\subfloat[Cumulative infections over time.\label{fig:b_cumulative_infections_model}]{
    \includegraphics[width=0.3\linewidth]{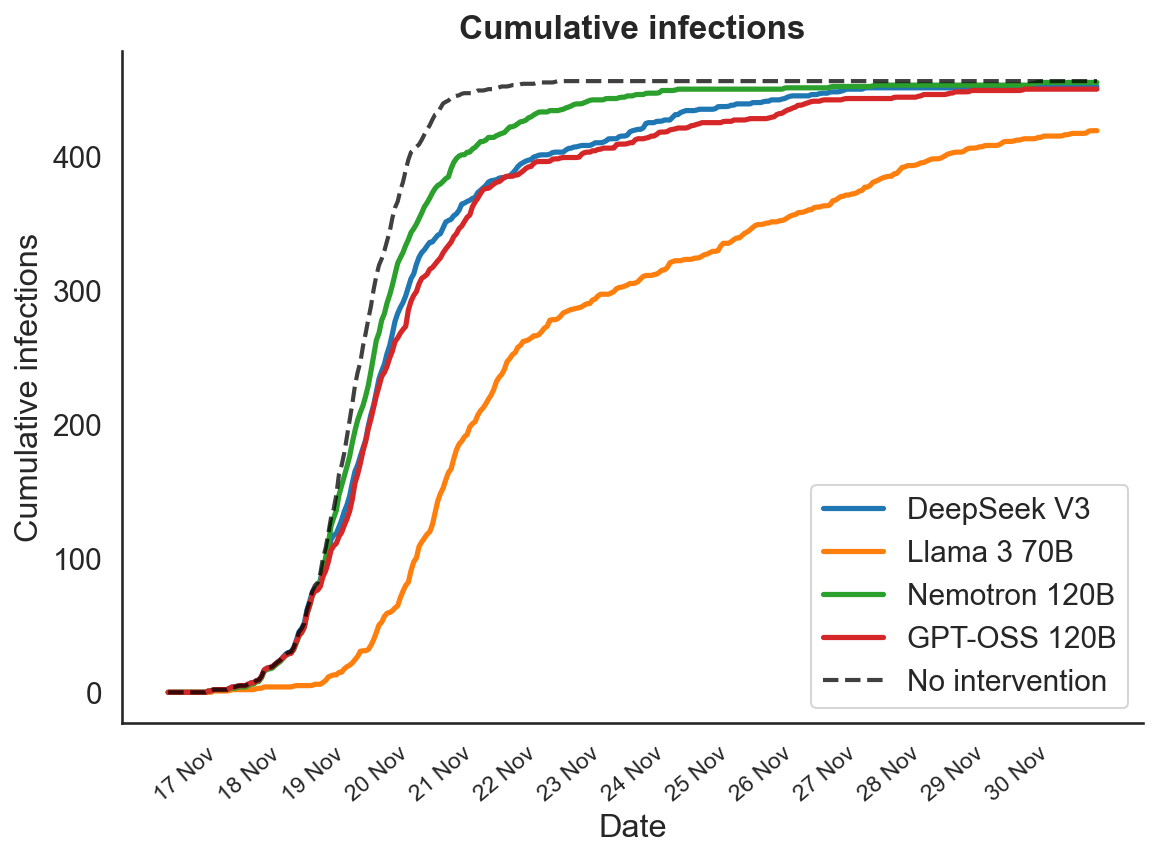}
}
\hfill
\subfloat[Daily quarantine rate.\label{fig:c_quarantine_rate_model}]{
    \includegraphics[width=0.3\linewidth]{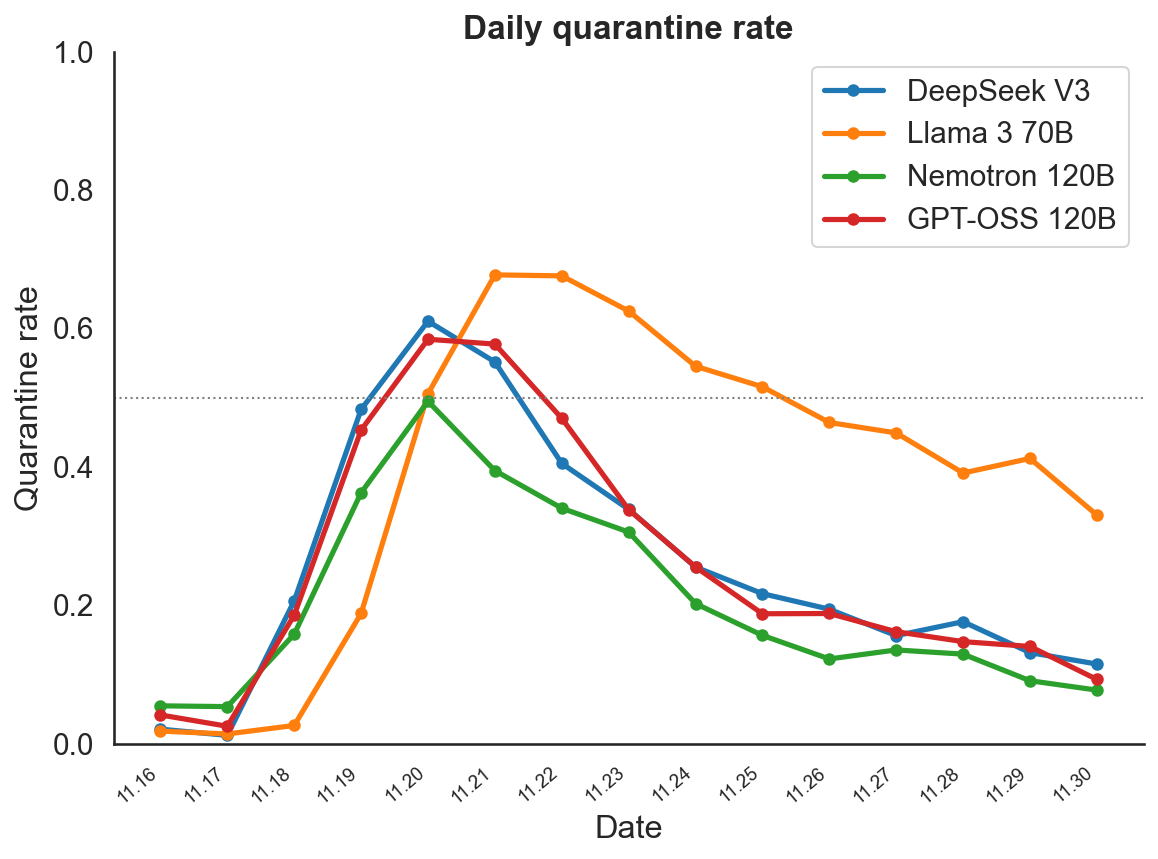}
}

\caption{Epidemic dynamics and quarantine behaviour across LLM architectures, with the ``no intervention'' reference shown as dashed lines. Panels show A) shows the number of active infections over time, B) cumulative infections, and C) the mean daily quarantine rate chosen by agents. All intervention conditions reduce and delay peak infections relative to the no-intervention baseline, but substantial behavioural variation emerges across architectures. Dense models exhibit divergent quarantine trajectories and epidemic outcomes, with Llama 3 70B sustaining higher quarantine rates for longer durations and consequently producing the lowest cumulative infection burden. In contrast, other models converge toward similar epidemic trajectories despite differing short-term quarantine responses.}
\label{fig:model_epidemic_dynmaics}
\end{figure*}

All LLMs, as seen in Figure~\ref{fig:a_active_infections_model}, produced lower epidemic peaks than the baseline, indicating that agents across all models engaged in quarantine behaviour to some extent. However, their underlying architecture influenced the effectiveness of the quarantine intervention and the resulting epidemic trajectories. Firstly, heterogeneous architectures produced different strategies for disease containment. We can see that Llama 3 70B resulted in a more delayed and flattened epidemic peak compared to Nemotron 120B, which exhibited higher peak active infections. This suggests that different architectures embody distinct behavioural priors, either biased toward quarantine compliance or toward maximising the game's objective function. Similarly, models sharing the same architectural family exhibited convergent behaviour. DeepSeek V3 and GPT-OSS 120B, both of which utilise a Mixture-of-Experts architecture, aligned closely in their quarantine rates and cumulative infection trajectories. 

\subsection{Effect of Geographic Framing on Epidemic Dynamics}
\label{results_geographic}

\begin{figure*}
\centering

\subfloat[Active infections over time. \label{fig:c_quarantine_rate_countries}]{
    \includegraphics[width=0.3\linewidth]{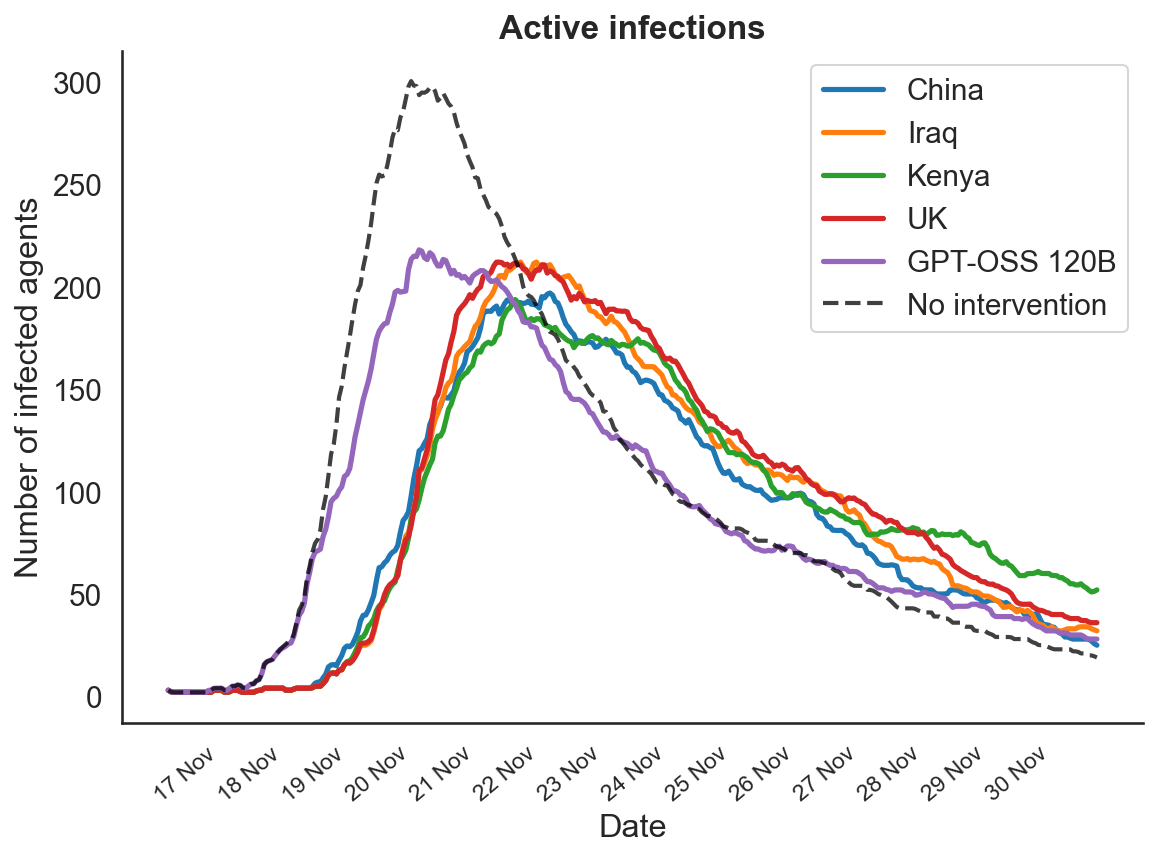}
}
\hfill
\subfloat[Cumulative infections over time.\label{fig:b_cumulative_infections_countries}]{
    \includegraphics[width=0.3\linewidth]{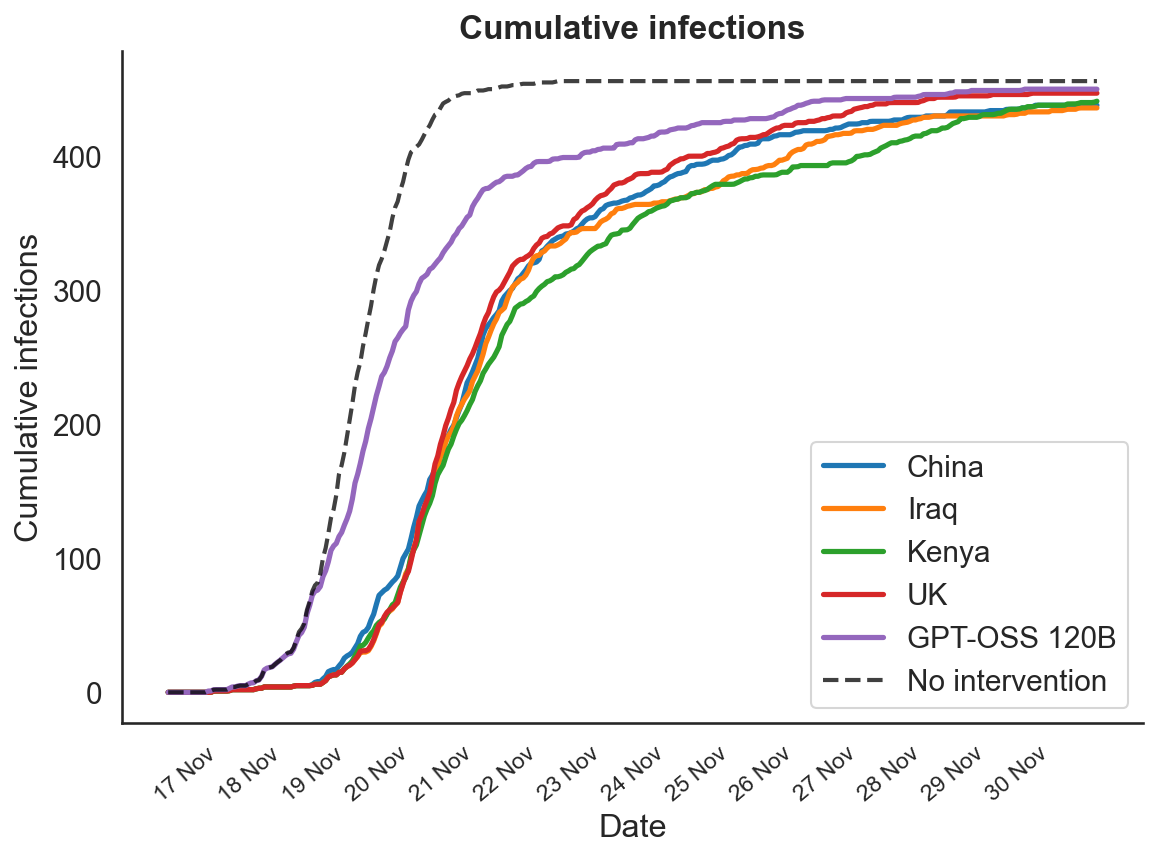}
}
\hfill
\subfloat[Daily quarantine rate.\label{fig:c_quarantine_rate_countries}]{
    \includegraphics[width=0.3\linewidth]{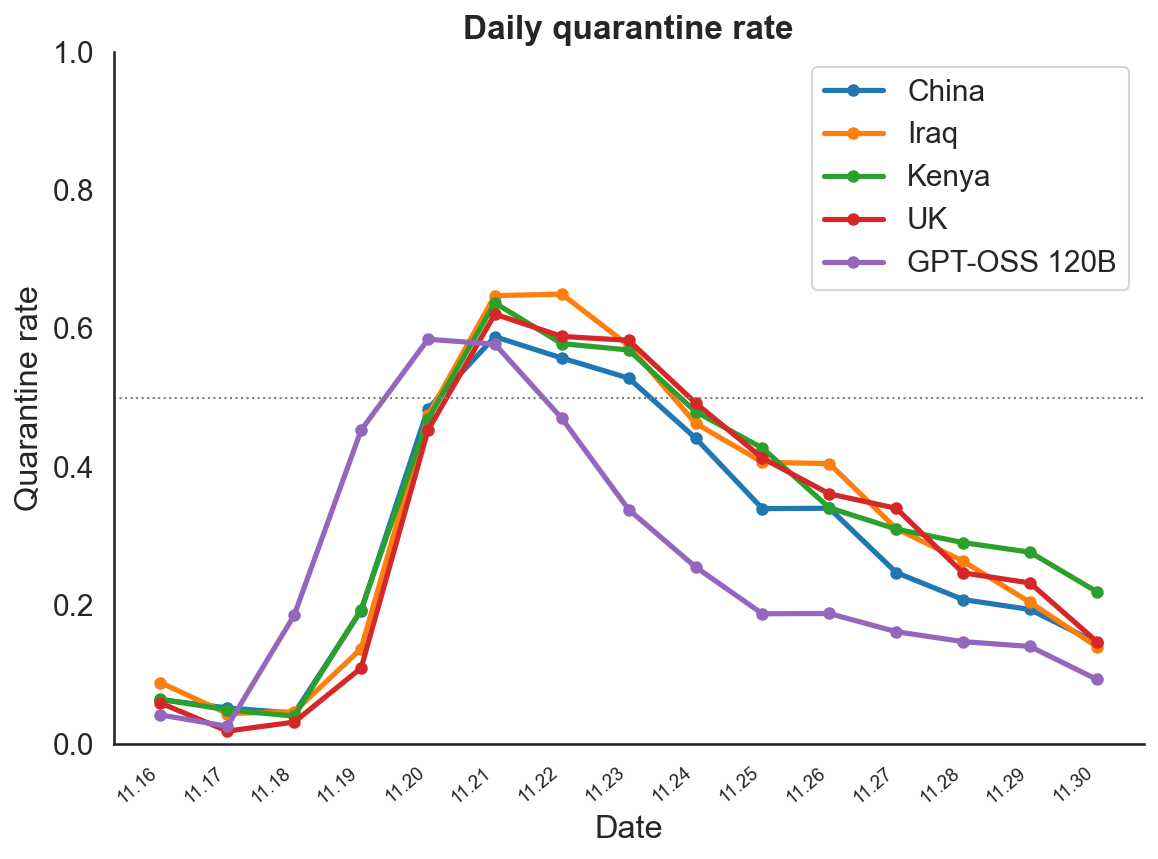}
}

\caption{Epidemic dynamics and quarantine behaviour across geographically-framed prompt conditions using the GPT-OSS 120B architecture. Agents were assigned one of four geographic identities (China, Iraq, Kenya, or the United Kingdom), which were included only as prompt framing variables while all epidemiological parameters remained fixed. Panels show A) active infections over time, B) cumulative infections, and C) the mean daily quarantine rate. Geographic framing produces measurable differences in quarantine uptake and epidemic progression despite identical underlying simulation conditions. The baseline GPT-OSS condition exhibits earlier and stronger quarantine adoption, resulting in lower infection peaks and cumulative infections, whereas geographically-framed agents display broadly similar but slightly diverging behavioural and epidemic trajectories over time.}
\label{fig:country_epidemic}
\end{figure*}

All four geographic conditions produced a notably delayed epidemic peak relative to both the ``no intervention'' baseline and the GPT-OSS 120B model (Figure \ref{fig:country_epidemic}). The ``no intervention'' curve peaks sharply around 21--22 November at approximately 300 active infections, whereas all geographically-framed conditions peak later and at a substantially lower magnitude of around 200 active infections.
Quarantine compliance was uniformly low across all geographic conditions during the first three simulation days (17--19 November), with mean daily rates ranging from 3.5\% (UK) to 5.8\% (Iraq) and no condition exceeding 8.5\% on any individual day. Despite this slow initial uptake, compliance rose sharply around day 4--5, peaking on day 6 across all conditions at rates between 58.5\% and 64.6\%, before declining steadily through the remainder of the simulation. The trajectories were structurally identical across all four conditions: a slow initial build, a sharp rise, a single peak, and a gradual decline. The cumulative infection panel reflects this pattern, with the delayed onset suggesting that initially infected agents prioritised isolation and reduced transmission, even at the expense of their own game objective.
Despite differences in geographic label, the four conditions converged rapidly in their epidemic trajectories. By the final simulation day, cumulative infections ranged narrowly from 436 (Iraq, attack rate 95.6\%) to 447 (United Kingdom, attack rate 98.0\%), a difference of just 11 agents across conditions. Kruskal-Wallis testing on the full time series of daily quarantine rates revealed no significant difference between conditions ($H=0.22,\; p=0.97$), and all six pairwise Mann-Whitney U tests were non-significant after Bonferroni correction. Pearson correlations between pairwise cumulative infection trajectories exceeded r=0.997 in every case (all $p < 10^{-16}$), confirming near-perfect convergence across geographic labels.
Notably, the GPT-OSS 120B unframed baseline diverged from all four geographic conditions, exhibiting more erratic quarantine behaviour with a lower and less sustained compliance peak, corresponding to its earlier and higher infection peak in the active infections panel. 

\subsection{Behavioural Replication and Health Belief as Motivation}
\label{results_behavioural_replication}

The binomial GLM achieved a Cox--Snell pseudo R\textsuperscript{2} of 0.055. Although modest, this value is consistent with the noisy behavioural signal observed in the corresponding human trial (pseudo R\textsuperscript{2} = 0.072) and is in the range typically observed in models of complex social and health-related decision making \citep{salecha2024large, gallegos-etal-2024-bias}. The full GLM results are presented in Table \ref{tab:binomial_glm}.

Perceived health severity emerged as the strongest statistical driver of quarantine behaviour (coefficient = +0.331, p = 0.002), indicating that higher perceived severity substantially increased the log-odds of quarantining. Perceived infection risk showed marginal significance (coefficient = +0.200, p = 0.076). In contrast, quarantine self-efficacy and response efficacy exhibited weak associations in this additive model.

\begin{table}[htbp]
\begin{center}
\caption{Binomial generalized linear model predicting quarantine behaviour.}
\begin{tabular}{lccc}
\toprule
Predictor & Coefficient (95\% CI) & SE & $p$ \\
\midrule
Intercept & $-1.773$ [$-2.757$, $-0.788$] & 0.502 & $<.001$ \\
Group (B vs.\ A) & $-0.178$ [$-0.642$, $0.287$] & 0.237 & .453 \\
Perceived infection risk & $0.200$ [$-0.021$, $0.421$] & 0.113 & .076 \\
Perceived health severity & $0.331$ [$0.124$, $0.538$] & 0.106 & .002 \\
Quarantine self-efficacy & $-0.062$ [$-0.268$, $0.144$] & 0.105 & .556 \\
Quarantine response efficacy & $0.025$ [$-0.213$, $0.263$] & 0.121 & .839 \\
\botrule
\end{tabular}
\label{tab:binomial_glm}
\end{center}
\end{table}


The resulting parameters of the Poisson regression model are presented in Table~\ref{tab:poisson_glm_interactions}. Agents in the low-cost group with low scores on the questionnaire predictors were taken to be the baseline characterised by the intercept. Relative to this, agents in the high-cost group had a significantly lower quarantine rate, indicating that higher quarantine costs reduced quarantine behaviour among agents with low motivation to quarantine ($p = 0.013$). Perceived infection risk and health severity were both positively associated with quarantine rates, indicating that agents were more likely to quarantine when they perceived infection as more likely/severe. These effects did not differ significantly by group, as neither interaction with group membership was statistically significant. Quarantine self-efficacy showed a significant negative association with quarantine rates in the low-cost group, although this effect was again not significantly moderated by group membership. Quarantine response efficacy was not significantly associated with quarantine behaviour, and its interaction with group membership was also non-significant.


\begin{table*}[ht]
\begin{center}
\caption{Poisson generalized linear model predicting quarantine behaviour with group interactions.}
\begin{tabular}{lccc}
\toprule
Predictor & Coefficient (95\% CI) & SE & $p$ \\
\midrule
Intercept & $-1.722$ [$-2.006$, $-1.439$] & 0.145 & $<.001$ \\
Group (B vs.\ A) & $-0.543$ [$-0.969$, $-0.117$] & 0.217 & .013 \\
Perceived infection risk & $0.148$ [$0.084$, $0.213$] & 0.033 & $<.001$ \\
Perceived infection risk $\times$ Group (B vs.\ A) & $-0.017$ [$-0.115$, $0.082$] & 0.050 & .739 \\
Perceived health severity & $0.194$ [$0.129$, $0.259$] & 0.033 & $<.001$ \\
Perceived health severity $\times$ Group (B vs.\ A) & $0.055$ [$-0.040$, $0.150$] & 0.048 & .256 \\
Quarantine self-efficacy & $-0.075$ [$-0.145$, $-0.006$] & 0.036 & .034 \\
Quarantine self-efficacy $\times$ Group (B vs.\ A) & $0.086$ [$-0.011$, $0.183$] & 0.049 & .083 \\
Quarantine response efficacy & $0.021$ [$-0.058$, $0.100$] & 0.040 & .604 \\
Quarantine response efficacy $\times$ Group (B vs.\ A) & $0.001$ [$-0.112$, $0.113$] & 0.057 & .991 \\
\botrule
\end{tabular}

\label{tab:poisson_glm_interactions}
\end{center}
\end{table*}

\subsection{Behavioural Response to Alternative Incentive Structures}
\label{results_incentive_structures}

\begin{figure*}
\centering

\subfloat[Active infections over time. \label{fig:a_active_infections_pointsystem}]{
    \includegraphics[width=0.3\linewidth]{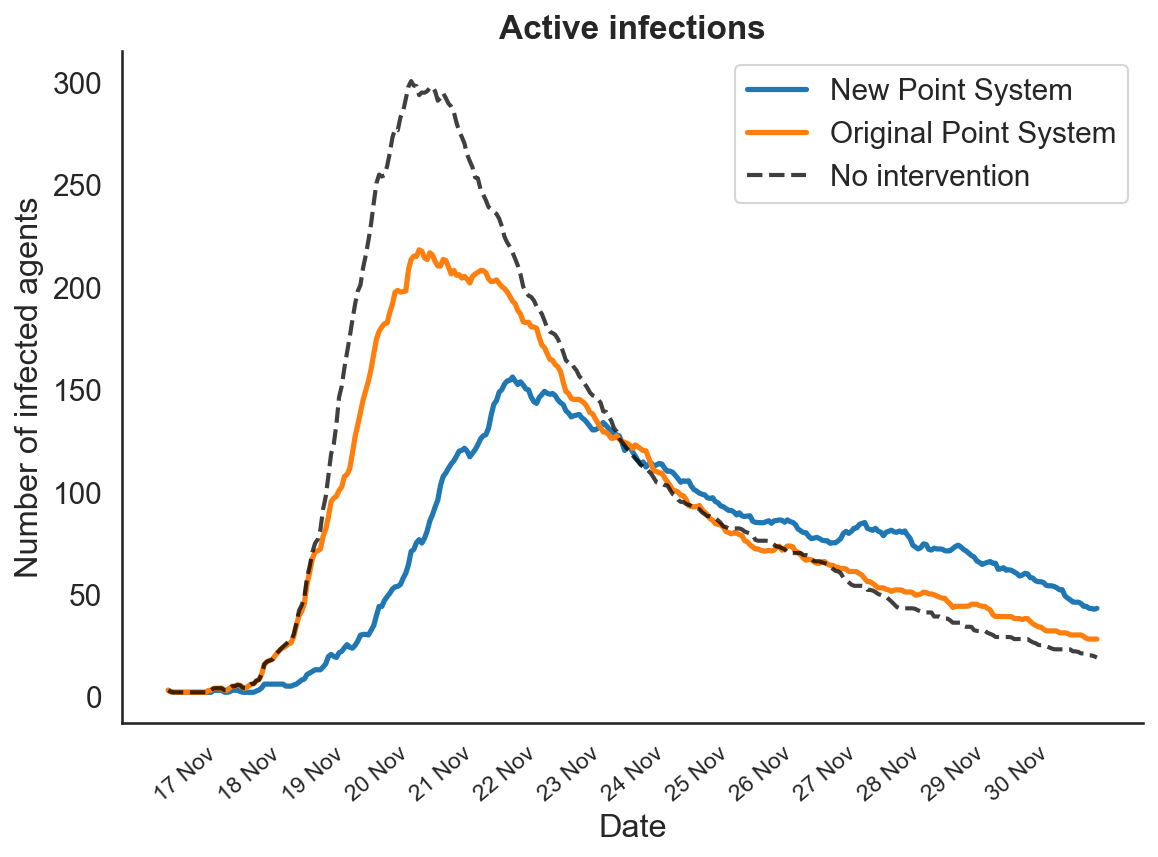}
}
\hfill
\subfloat[Cumulative infections over time.\label{fig:b_cumulative_infections_pointsystem}]{
    \includegraphics[width=0.3\linewidth]{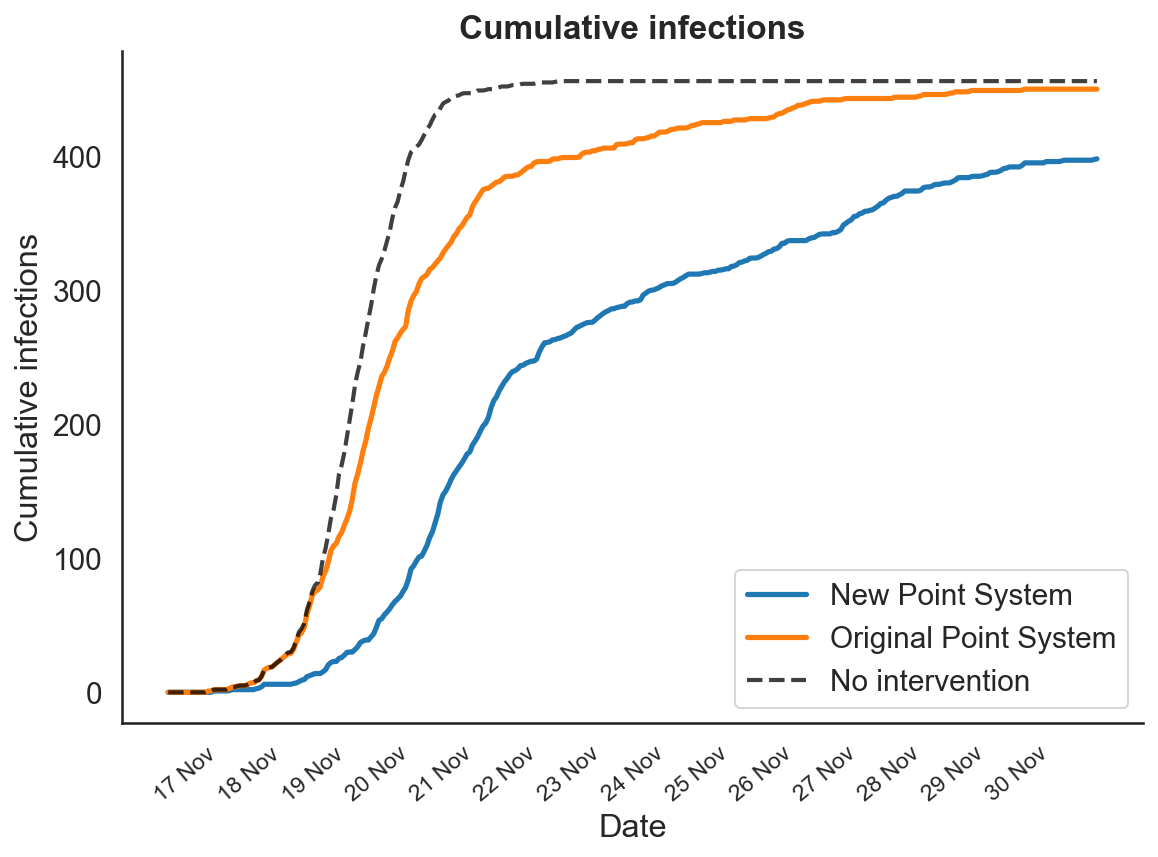}
}
\hfill
\subfloat[Daily quarantine rate.\label{fig:c_quarantine_rate_pointsystem}]{
    \includegraphics[width=0.3\linewidth]{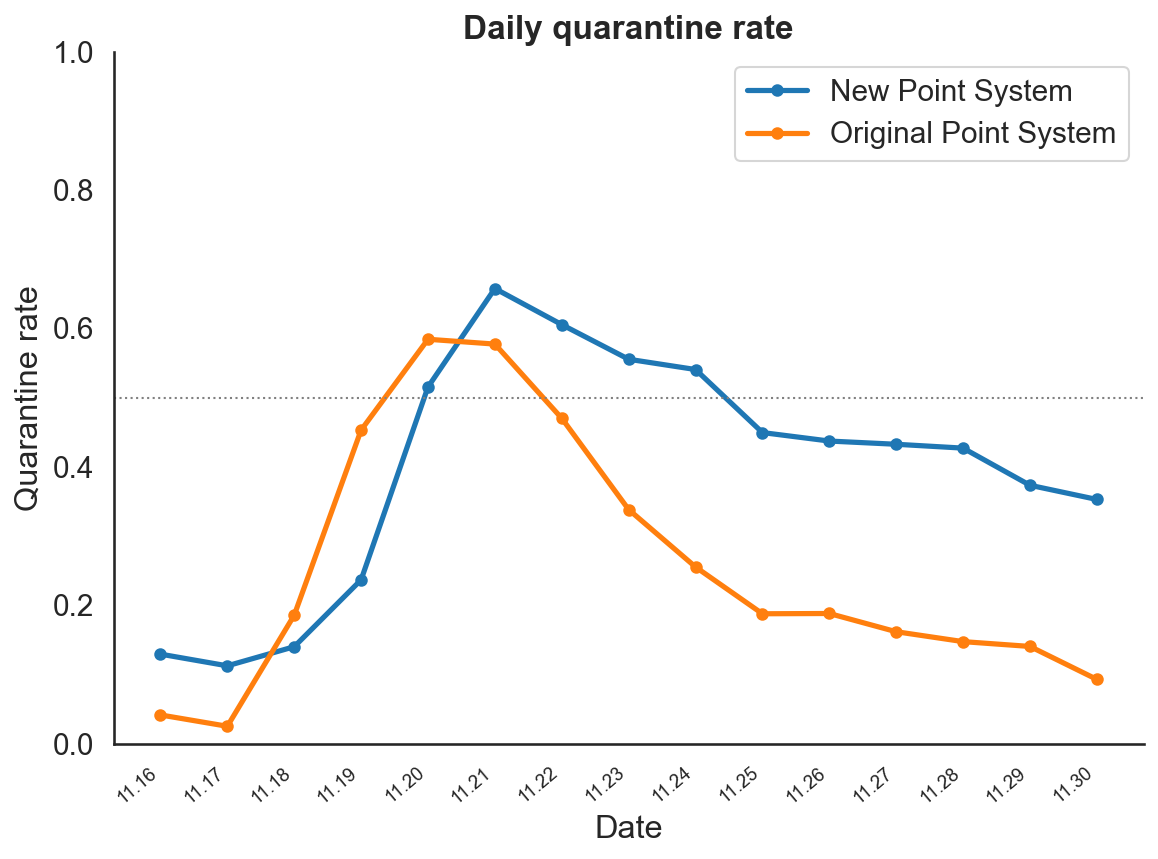}
}

\caption{Epidemic dynamics and quarantine behaviour under the new point system compared to the original point system, with the ``no intervention'' reference shown as dashed lines. The revised incentive structure reduces peak prevalence from ~0.5 to ~0.3 and prolongs compliance, suggesting that reward design has a meaningful impact.}
\label{fig:degree_comparison}
\end{figure*}

Alternative incentives within the framework led to different epidemic outcomes. The new points system notably reduced the number of infections (Figure~\ref{fig:b_cumulative_infections_pointsystem}) and flattened the epidemic curve (Figure~\ref{fig:a_active_infections_pointsystem}), lowering peak prevalence from $\sim0.5$ under the original point system to $\sim0.3$. Although agents were initially slower to quarantine, overall quarantine uptake eventually reached similar levels in both simulations ($\sim0.6$-$0.7$; Figure~\ref{fig:c_quarantine_rate_pointsystem}). Under the revised system, however, agents maintained quarantine behaviour for longer, with substantially higher compliance persisting until the end of the simulation.

    
    

\subsection{Behavioural Consistency of LLM-Driven Agents}
\label{consitency_section}

\begin{figure*}
    \centering
    \includegraphics[width=\linewidth]{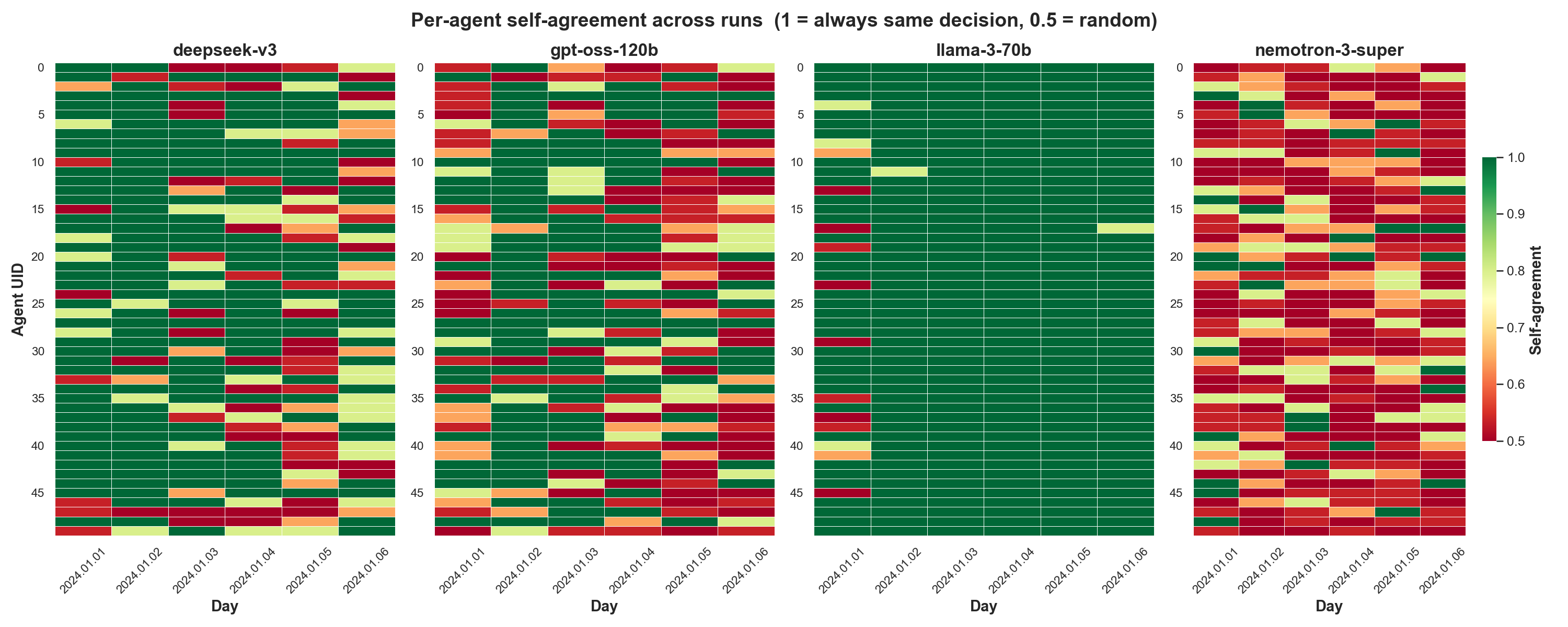}
    \caption{Self-agreement heatmap across LLM architectures across 10 runs. Each cell represents the self-agreement score between repeated simulation runs for a given agent under identical prompt conditions for a given day, with scores ranging from 0.5 (fully stochastic) to 1 (fully deterministic). Llama 3 70B shows near-perfect consistency, while Nemotron 120B exhibits substantial stochasticity; this distinction has implications for whether observed epidemic trajectory differences across architectures reflect genuine behavioural priors or seed-level noise.}
    \label{fig:selfagreement}
\end{figure*}
High-consistency architectures offer greater internal validity for testing specific behavioural rules, while high-variance models may more faithfully represent the decision-making of real human populations. As seen in Figure \ref{fig:selfagreement}, Llama 3 70B exhibited the highest self-agreement, with most scores being $\mathcal{C} = 1$, indicating consistent agent decisions across all runs. DeepSeek V3 and GPT-OSS 120B showed moderate consistency ($\mathcal{C} \in [0.7, 0.9]$), while Nemotron 120B was the most stochastic, with scores around $\mathcal{C} \approx 0.6$. 

\section{Discussion}



In this paper we have introduced the Epi-LLM framework and explored how it can be used to understand spread of an outbreak across a contact network. We show that architectural design choice is a vital consideration when developing LLM-driven agent-based epidemic simulations. Different architectures and geographic context provided to the agents exhibit distinct decision-making behaviours, and the stochasticity and consistency of these models varies across architectures. This may be either desirable or not, depending on the simulation objective. If the goal is to control agent behaviour, low-variance architectures are preferable, whereas if the goal is to emulate the inherent unpredictability of real-world human decision-making, higher variance models may be more appropriate.

The geographic framing experiment found no significant differences in either quarantine behaviour or epidemic outcomes across national conditions, suggesting that geographic labels alone do not induce culturally differentiated behaviour in LLM agents. Although real-world compliance behaviour varies systematically across cultural contexts \citep{chen2021culture, rajkumar2021relationship}, the prompts in this study encoded only as a transmission-risk modifier, without incorporating culturally grounded attitudes such as institutional trust or collective responsibility. The results therefore suggest that meaningful geographic variation in GABMs likely requires explicit behavioural parameterisation.
This null result is ambiguous: it may reflect genuine LLM insensitivity to national context, or insufficient cultual content in the prompt.

The findings suggest that LLM agents can reproduce broad features of health-belief-driven quarantine behaviour, particularly the tendency for higher perceived infection risk and severity to increase compliance. Economic barriers also reduced quarantine behaviour, partially replicating the filtering effect observed in the AUIB study, although agents did not reproduce the more nuanced human pattern in which stronger health beliefs moderated the impact of quarantine cost. The alternative incentive experiment further demonstrated that modifying reward structures alone can substantially alter epidemic trajectories, highlighting the framework's potential as an in silico environment for testing intervention strategies and policy designs prior to real-world deployment.


This is the first work to integrate large language models, agent-based epidemic modelling, and a real-world behavioural epi-game into a unified simulation framework. Prior GABMs have shown that LLM-driven agents produce plausible social behaviours \citep{park2023generative, choi2025infected, williams2023epidemic} but have relied on a single architecture, small populations, and limited empirical grounding. The Epi-LLM framework addresses each of these gaps. The most substantive novel contribution is the use of real psychosocial survey data from the AUIB epi-game to parameterise agent belief profiles, mapping Health Belief Model responses directly onto agent attributes and transforming a static behavioural dataset into a dynamic simulation input. A further contribution is the systematic comparison of four LLM architectures under identical conditions, showing that architecture shapes emergent population-level behaviour — with implications for the design of synthetic societies. The self-agreement analysis provides a reusable tool for characterising agent stochasticity prior to deployment, and the incentive structure experiment demonstrates the framework's utility as a prospective sandbox for testing epi-game design choices and public health intervention scenarios.


This work has some limitations. Firstly, each LLM architecture was evaluated in a single simulation run with a fixed random seed (Section~\ref{llmarchitectyre_section}). Although the architecture comparison reveals clear differences in quarantine trajectories and epidemic outcomes, without replication across seeds it is not possible to formally separate architectural effects from within-model stochastic variation. The self-agreement analysis (Section~\ref{consitency_section}) reinforces this concern: high-variance architectures such as Nemotron 120B exhibited self-agreement scores around 0.6, meaning a non-trivial proportion of decisions would differ under identical conditions.

Secondly, the null result in the geographic framing experiment (Section~\ref{results_geographic}) reflects the specific way in which national context was operationalised: as a physical exposure modifier rather than as a set of culturally grounded attitudes. The prompt contained no information about compliance norms, institutional trust, or the social cost of non-compliance. We therefore cannot distinguish between two interpretations: that LLMs are genuinely insensitive to national context, or that the prompt design did not provide sufficient cultural content to elicit differentiated behaviour. Resolving this distinction requires a revised experiment with explicit attitudinal parameterisation.

Thirdly, agents reproduced threat-driven quarantine patterns but not the finer moderation of quarantine cost by health-belief scores observed in the human trial. This partial replication may reflect alignment bias, where RLHF-trained models favour normative responses such as quarantining under high prevalence \citep{salecha2024large, gallegos-etal-2024-bias}, or task inference, where agents infer the objective of minimising infection — both of which could produce broad threat-driven compliance without nuanced cost-moderation effects. The GLMs also do not account for repeated observations or network dependencies; mixed-effects or network-aware models would strengthen these conclusions.

Fourthly, the alternative incentive structure (Section~\ref{results_incentive_structures}) was evaluated under a single architecture (GPT-OSS 120B). Given that architecture is a determinant of quarantine behaviour, the degree to which the observed response to the redesigned reward function generalises across other models remains untested.


Several extensions could strengthen both the methodological robustness and empirical relevance of this framework. Future work should assess sensitivity to prompt design through systematic pertubations. The contact network could be refined to incorporate long-term contacts and community clustering. Model biases warrant investigation across demographic and cultural framings, and introducting multi-objective rewards, repeated seeding, and richer epidemiological indicatiors would improve calibration and realism

Finally, expanding the information available to agents, such as aggregate epidemiological indicators such as the reproduction number, would better reflect real-world decision environments. 

\section{Conclusion}

Taken together, these results establish that the Epi-LLM framework is technically feasible, empirically grounded, and capable of generating substantive insights about behavioural dynamics in epidemic settings. The pseudo-$R^2$ values are consistent with those from the human trial and with the broader literature on complex social decision-making. LLM architecture proved a non-trivial determinant of emergent behaviour, with implications for how synthetic societies should be designed depending on whether the goal is controlled hypothesis testing or realistic heterogeneity. The geographic framing null result is informative: culturally realistic agents will require explicit attitudinal parameterisation rather than demographic labelling alone. The incentive-structure experiment illustrated a practical use case as a prospective design tool for testing epi-game mechanics \textit{in silico} before real-world deployment. Whether the framework can serve as a quantitatively accurate model of a specific human population rather than a qualitatively plausible one requires formal validation, which we regard as the natural next step. The present work lays the methodological foundation: it demonstrates what prompting strategies can and cannot achieve, where architectural choices matter, and how real-world survey data can be embedded into generative agent behaviour.


\section{Conflicts of interest}
The authors declare that they have no competing interests.

\section{Funding}

P.F., A.K., T.O., and L.S. are all supported by the EPSRC Centre for Doctoral Training in Healthcare Data Science (EP/Y035321/1). A.K is also supported in part by the Leverhulme Trust (Grant RC-2018-003) for the Leverhulme Centre for Demographic Science.

\section{Data availability}
The data underlying this article are available upon request at \href{https://doi.org/10.5281/zenodo.18209232}{doi.org/10.5281/zenodo.18209232}

\section{Author contributions statement}

J.P.G. and A.C. came up with the concept of combining epigames and agent-based models. P.F., A.K., T.O. and L.S. extended the concept to include LLMs. P.F., A.K., T.O. and L.S. all contributed equally to the design of the methods, the analysis and the generation of the results. P.F., A.K., T.O. and L.S. drafted the initial manuscript, with support from A.C., F.d.L. and J.P.G.. All authors approved the final manuscript.

\section{Acknowledgments}

The authors would like to thank Luca Ferretti and Charles Rahal for their invaluable comments on the manuscript.

\bibliographystyle{oup-abbrvnat-custom}
\bibliography{sample}

@inproceedings{park2023generative,
  title={Generative agents: Interactive simulacra of human behavior},
  author={Park, Joon Sung and O'Brien, Joseph and Cai, Carrie Jun and Morris, Meredith Ringel and Liang, Percy and Bernstein, Michael S},
  booktitle={Proceedings of the 36th annual acm symposium on user interface software and technology},
  pages={1--22},
  year={2023}
}

@article{colubri2026understanding,
  title={Understanding human behaviour for pandemic preparedness with epigames},
  author={Colubri, Andr{\'e}s and Williams, Dmitri and Valente, Thomas and Bauch, Chris T and Drake, John M and Mills, Melinda C and Drury, John and Fraser, Christophe and Ferretti, Luca and Panovska-Griffiths, Jasmina},
  journal={Nature Health},
  pages={1--3},
  year={2026},
  publisher={Nature Publishing Group UK London}
}

@article{liu2024deepseek,
  title={Deepseek-v3 technical report},
  author={Liu, Aixin and Feng, Bei and Xue, Bing and Wang, Bingxuan and Wu, Bochao and Lu, Chengda and Zhao, Chenggang and Deng, Chengqi and Zhang, Chenyu and Ruan, Chong and others},
  journal={arXiv preprint arXiv:2412.19437},
  year={2024}
}

@article{grattafiori2024llama,
  title={The llama 3 herd of models},
  author={Grattafiori, Aaron and Dubey, Abhimanyu and Jauhri, Abhinav and Pandey, Abhinav and Kadian, Abhishek and Al-Dahle, Ahmad and Letman, Aiesha and Mathur, Akhil and Schelten, Alan and Vaughan, Alex and others},
  journal={arXiv preprint arXiv:2407.21783},
  year={2024}
}

@article{blakeman2025nvidia,
  title={NVIDIA Nemotron 3: Efficient and Open Intelligence},
  author={Blakeman, Aaron and Grattafiori, Aaron and Basant, Aarti and Gupta, Abhibha and Khattar, Abhinav and Renduchintala, Adi and Vavre, Aditya and Shukla, Akanksha and Bercovich, Akhiad and Ficek, Aleksander and others},
  journal={arXiv preprint arXiv:2512.20856},
  year={2025}
}

@article{agarwal2025gpt,
  title={gpt-oss-120b \& gpt-oss-20b model card},
  author={Agarwal, Sandhini and Ahmad, Lama and Ai, Jason and Altman, Sam and Applebaum, Andy and Arbus, Edwin and Arora, Rahul K and Bai, Yu and Baker, Bowen and Bao, Haiming and others},
  journal={arXiv preprint arXiv:2508.10925},
  year={2025}
}

@article{champion2008health,
  title={The health belief model},
  author={Champion, Victoria L and Skinner, Celette Sugg and others},
  journal={Health behavior and health education: Theory, research, and practice},
  volume={4},
  pages={45--65},
  year={2008}
}

@article{chen2021culture,
  title={Culture and contagion: Individualism and compliance with COVID-19 policy},
  author={Chen, Chinchih and Frey, Carl Benedikt and Presidente, Giorgio},
  journal={Journal of economic behavior \& organization},
  volume={190},
  pages={191--200},
  year={2021},
  publisher={Elsevier}
}

@article{rajkumar2021relationship,
  title={The relationship between measures of individualism and collectivism and the impact of COVID-19 across nations},
  author={Rajkumar, Ravi Philip},
  journal={Public Health in Practice},
  volume={2},
  pages={100143},
  year={2021},
  publisher={Elsevier}
}

@article{holme2012temporal,
  title={Temporal networks},
  author={Holme, Petter and Saram{\"a}ki, Jari},
  journal={Physics reports},
  volume={519},
  number={3},
  pages={97--125},
  year={2012},
  publisher={Elsevier}
}

@online{starsim2026,
  author  = {{Starsim}},
  title   = {Starsim: Agent-based disease modeling},
  year    = {2026},
  url     = {https://starsim.org/},
  urldate = {2026-04-20}
}

@online{openrouter2026,
  author  = {{OpenRouter}},
  title   = {OpenRouter API},
  year    = {2026},
  url     = {https://openrouter.ai/},
  urldate = {2026-04-21}
}

@article{kaur2025ai,
  title={AI-driven epidemic intelligence: the future of outbreak detection and response},
  author={Kaur, Jasleen and Butt, Zahid Ahmad},
  journal={Frontiers in Artificial Intelligence},
  volume={8},
  pages={1645467},
  year={2025},
  publisher={Frontiers Media SA}
}

@inproceedings{wong2025comparative,
  title={Comparative Evaluation and Performance of Large Language Models in Clinical Infection Control Scenarios: A Benchmark Study},
  author={Wong, Shuk-Ching and Chiu, Edwin Kwan-Yeung and Chiu, Kelvin Hei-Yeung and Tam, Anthony Raymond and Chau, Pui-Hing and Choi, Ming-Hong and Ng, Wing-Yan and Kwok, Monica Oi-Tung and Chau, Benny Yu and Ng, Michael Yuey-Zhun and others},
  booktitle={Healthcare},
  volume={13},
  number={20},
  pages={2652},
  year={2025},
  organization={MDPI}
}

@article{samaei2026epidemiqs,
  title={EpidemIQs: Prompt-to-paper LLM agents for epidemic modeling and analysis},
  author={Samaei, Mohammad Hosseini and Sahneh, Faryad Darabi and Cohnstaedt, Lee W and Scoglio, Caterina M},
  journal={IEEE Transactions on Artificial Intelligence},
  year={2026},
  publisher={IEEE}
}

@article{choi2025infected,
  title={Infected Smallville: How Disease Threat Shapes Sociality in LLM Agents},
  author={Choi, Soyeon and Lee, Kangwook and Sng, Oliver and Ackerman, Joshua M},
  journal={arXiv preprint arXiv:2506.13783},
  year={2025}
}

@article{ye2025integrating,
  title={Integrating artificial intelligence with mechanistic epidemiological modeling: a scoping review of opportunities and challenges},
  author={Ye, Yang and Pandey, Abhishek and Bawden, Carolyn and Sumsuzzman, Dewan Md and Rajput, Rimpi and Shoukat, Affan and Singer, Burton H and Moghadas, Seyed M and Galvani, Alison P},
  journal={Nature Communications},
  volume={16},
  number={1},
  pages={581},
  year={2025},
  publisher={Nature Publishing Group UK London}
}

@inproceedings{jian2017applying,
  title={Applying deep learning for surrogate construction of simulation systems},
  author={Jian, Zong-De and Chang, Hung-Jui and Hsu, Tsan-sheng and Wang, Da-Wei},
  booktitle={International Conference on Simulation and Modeling Methodologies, Technologies and Applications},
  pages={335--350},
  year={2017},
  organization={Springer}
}

@article{lu2024llms,
  title={LLMs and generative agent-based models for complex systems research},
  author={Lu, Yikang and Aleta, Alberto and Du, Chunpeng and Shi, Lei and Moreno, Yamir},
  journal={Physics of Life Reviews},
  volume={51},
  pages={283--293},
  year={2024},
  publisher={Elsevier}
}

@article{williams2023epidemic,
  title={Epidemic modeling with generative agents},
  author={Williams, Ross and Hosseinichimeh, Niyousha and Majumdar, Aritra and Ghaffarzadegan, Navid},
  journal={arXiv preprint arXiv:2307.04986},
  year={2023}
}

@inproceedings{chuang2024simulating,
  title={Simulating opinion dynamics with networks of llm-based agents},
  author={Chuang, Yun-Shiuan and Goyal, Agam and Harlalka, Nikunj and Suresh, Siddharth and Hawkins, Robert and Yang, Sijia and Shah, Dhavan and Hu, Junjie and Rogers, Timothy},
  booktitle={Findings of the association for computational linguistics: NAACL 2024},
  pages={3326--3346},
  year={2024}
}

@article{rizzo2024future,
  title={The future of large language models in fighting emerging outbreaks: lights and shadows},
  author={Rizzo, Alberto and Mensa, Enrico and Giacomelli, Andrea},
  journal={The Lancet Microbe},
  volume={5},
  number={11},
  year={2024},
  publisher={Elsevier}
}

@article{biswas2014seir,
  title={A SEIR model for control of infectious diseases with constraints},
  author={Biswas, Md Haider Ali and Paiva, Lu{\'\i}s Tiago and De Pinho, MDR},
  journal={Mathematical Biosciences and Engineering},
  volume={11},
  number={4},
  pages={761},
  year={2014},
  publisher={American Institute of Mathematical Sciences}
}

@article{weiss2013sir,
  title={The SIR model and the foundations of public health},
  author={Weiss, Howard Howie},
  journal={Materials matematics},
  pages={0001--17},
  year={2013}
}

@article{colubri2026app,
  title={App-based Epidemic Game to Model Belief-Behavior Mapping and Cost Incentives in Voluntary Quarantine: A Randomized Controlled Trial},
  author={Colubri, Andr{\'e}s and Grozdani, Andonaq and Khandpekar, Mansi and Graytee, Yousif and Al-Mohammedi, Omar and Al-Shabandar, Ahmed Ayden and Shabeeb, Wid Yasir and Ghassan, Yaqoob and Swayedi, Hayder and Bauch, Chris T and others},
  journal={medRxiv},
  pages={2026--01},
  year={2026},
  publisher={Cold Spring Harbor Laboratory Press}
}

@article{yang2024swe,
  title={Swe-agent: Agent-computer interfaces enable automated software engineering},
  author={Yang, John and Jimenez, Carlos E and Wettig, Alexander and Lieret, Kilian and Yao, Shunyu and Narasimhan, Karthik and Press, Ofir},
  journal={Advances in Neural Information Processing Systems},
  volume={37},
  pages={50528--50652},
  year={2024}
}

@article{lu2024ai,
  title={The ai scientist: Towards fully automated open-ended scientific discovery},
  author={Lu, Chris and Lu, Cong and Lange, Robert Tjarko and Foerster, Jakob and Clune, Jeff and Ha, David},
  journal={arXiv preprint arXiv:2408.06292},
  year={2024}
}

@article{boiko2023emergent,
  title={Emergent autonomous scientific research capabilities of large language models},
  author={Boiko, Daniil A and MacKnight, Robert and Gomes, Gabe},
  journal={arXiv preprint arXiv:2304.05332},
  year={2023}
}

@article{tracy2018agent,
  title={Agent-based modeling in public health: current applications and future directions},
  author={Tracy, Melissa and Cerd{\'a}, Magdalena and Keyes, Katherine M},
  journal={Annual review of public health},
  volume={39},
  pages={77--94},
  year={2018},
  publisher={Annual Reviews}
}

@article{kermack1927contribution,
  title={A contribution to the mathematical theory of epidemics},
  author={Kermack, William Ogilvy and McKendrick, Anderson G},
  journal={Proceedings of the royal society of london. Series A, Containing papers of a mathematical and physical character},
  volume={115},
  number={772},
  pages={700--721},
  year={1927},
  publisher={The Royal Society London}
}

@article{park2024generative,
  title={Generative agent simulations of 1,000 people},
  author={Park, Joon Sung and Zou, Carolyn Q and Shaw, Aaron and Hill, Benjamin Mako and Cai, Carrie and Morris, Meredith Ringel and Willer, Robb and Liang, Percy and Bernstein, Michael S},
  journal={arXiv preprint arXiv:2411.10109},
  year={2024}
}

@article{gallegos-etal-2024-bias,
    title = "Bias and Fairness in Large Language Models: A Survey",
    author = "Gallegos, Isabel O.  and
      Rossi, Ryan A.  and
      Barrow, Joe  and
      Tanjim, Md Mehrab  and
      Kim, Sungchul  and
      Dernoncourt, Franck  and
      Yu, Tong  and
      Zhang, Ruiyi  and
      Ahmed, Nesreen K.",
    journal = "Computational Linguistics",
    volume = "50",
    number = "3",
    month = sep,
    year = "2024",
    address = "Cambridge, MA",
    publisher = "MIT Press",
    url = "https://aclanthology.org/2024.cl-3.8/",
    doi = "10.1162/coli_a_00524",
    pages = "1097--1179"
}

@article{salecha2024large,
  title={Large language models display human-like social desirability biases in Big Five personality surveys},
  author={Salecha, Aadesh and Ireland, Molly E and Subrahmanya, Shashanka and Sedoc, Jo{\~a}o and Ungar, Lyle H and Eichstaedt, Johannes C},
  journal={PNAS nexus},
  volume={3},
  number={12},
  pages={pgae533},
  year={2024},
  publisher={Oxford University Press US}
}

@article{kerr2021covasim,
  title={Covasim: an agent-based model of COVID-19 dynamics and interventions},
  author={Kerr, Cliff C and Stuart, Robyn M and Mistry, Dina and Abeysuriya, Romesh G and Rosenfeld, Katherine and Hart, Gregory R and N{\'u}{\~n}ez, Rafael C and Cohen, Jamie A and Selvaraj, Prashanth and Hagedorn, Brittany and others},
  journal={PLoS computational biology},
  volume={17},
  number={7},
  pages={e1009149},
  year={2021},
  publisher={Public Library of Science San Francisco, CA USA}
}

@article{danon2012social,
  title={Social encounter networks: collective properties and disease transmission},
  author={Danon, Leon and House, Thomas A and Read, Jonathan M and Keeling, Matt J},
  journal={Journal of The Royal Society Interface},
  volume={9},
  number={76},
  pages={2826--2833},
  year={2012},
  publisher={The Royal Society}
}

@book{ferguson2020report,
  title={Report 9: Impact of non-pharmaceutical interventions (NPIs) to reduce COVID19 mortality and healthcare demand},
  author={Ferguson, Neil M and Laydon, Daniel and Nedjati-Gilani, Gemma and Imai, Natsuko and Ainslie, Kylie and Baguelin, Marc and Bhatia, Sangeeta and Boonyasiri, Adhiratha and Cucunub{\'a}, Zulma and Cuomo-Dannenburg, Gina and others},
  volume={16},
  year={2020},
  publisher={Imperial College London London}
}

@article{ferguson2006strategies,
  title={Strategies for mitigating an influenza pandemic},
  author={Ferguson, Neil M and Cummings, Derek AT and Fraser, Christophe and Cajka, James C and Cooley, Philip C and Burke, Donald S},
  journal={Nature},
  volume={442},
  number={7101},
  pages={448--452},
  year={2006},
  publisher={Nature Publishing Group UK London}
}

@article{li2024agent,
  title={Agent hospital: A simulacrum of hospital with evolvable medical agents},
  author={Li, Junkai and Lai, Yunghwei and Li, Weitao and Ren, Jingyi and Zhang, Meng and Kang, Xinhui and Wang, Siyu and Li, Peng and Zhang, Ya-Qin and Ma, Weizhi and others},
  journal={arXiv preprint arXiv:2405.02957},
  year={2024}
}


\begin{appendices}
\twocolumn

\section{Prompt Engineering Structure}\label{app:Prompt Engineering Strucutre}

Prompt engineering has been widely employed to instantiate LLM agents in task-specific environments, providing the context necessary for coherent and goal-directed behavior \citep{park2023generative} in a cost effective manner. In this work, we designed a prompt that served as a comprehensive decision-support environment, equipping each agent with several layers of contextually relevant information.

Firstly, the agents were informed of the core mechanics and goals of the AUIB study \citep{colubri2026understanding}: maximising their accumulated point balance while avoiding infection. This framing ensured agents operated under the same incentive structure as human participants. To mirror the information available to real-world players, each agent was provided with its current health status (healthy, mildly symptomatic, or severely symptomatic), full infection history, current point balance, group allocation, and pre-assigned health beliefs. Moreover, to simulate realistic situational awareness, agents were also given network-level information: the prevalence of symptomatic individuals among their direct contacts, and the number of individuals in their immediate proximity currently under quarantine. The baseline prompt can be found in \ref{app:prompt_template}.
\onecolumn
\subsection{Prompt Templates}

\label{app:prompt_template}

\begin{tcolorbox}[
    colback=gray!5,
    colframe=gray!50,
    title=\textbf{Base Agent Prompt},
    fonttitle=\small,
    breakable,
    listing only,
    listing options={
        basicstyle=\ttfamily\small,
        breaklines=true,
        columns=fullflexible
    }
]
\begin{verbatim}

You are playing an epidemic decision game where your goal is to maximise your 
total points.

Game mechanics:
- A disease spreads through a contact network: interacting with others exposes 
you to infection risk.
- Your infection risk increases with local prevalence and your contacts.
- If infected, you may lose points (reduced rewards, possible large penalties).
- You move through health states (susceptible -> infected -> recovered).

Decision each round:
- Quarantine: {low_reward} pts. No infection risk this round.
- Stay active: {high_reward} pts. Risk infection from contacts.

This is a trade-off between:
- Short-term reward (staying active)
- Long-term risk (infection causing point losses)

Your objective:
Maximise your total points over time. Consider expected future losses from 
infection, not just immediate reward.

Your initial beliefs (from pregame survey, scale 1-6, where 1 = weakest, 
6 = strongest):
- Perceived infection risk: how likely you think it is that you will get 
infected.
- Perceived health severity: how serious you think infection would be for 
your health.
- Quarantine self-efficacy: how confident you are that you can successfully 
follow quarantine.
- Quarantine response efficacy: how effective you think quarantine is at 
preventing spread.

Local prevalence (0-1): fraction of contacts infected in the previous timestep.

Your current state:
- Time: {t}
- Status: {status}
- Infection history: {has_been_infected}
- Points: {points}
- Local prevalence: {local_prev}
- Perceived infection risk (1-6): {perceived_infection_risk}
- Perceived health severity (1-6): {perceived_health_severity}
- Quarantine self-efficacy (1-6): {quarantine_self_efficacy}
- Quarantine response efficacy (1-6): {quarantine_response_efficacy}

Use this framework to guide your decision. 
Should you quarantine this round? Reply with only 'yes' or 'no'.
\end{verbatim}
\end{tcolorbox}
\label{app:prompt_template_geography}

\begin{tcolorbox}[
    colback=gray!5,
    colframe=gray!50,
    title=\textbf{Augmented Prompt Template with Geographic Location},
    fonttitle=\small,
    breakable,
    listing only,
    listing options={
        basicstyle=\ttfamily\small,
        breaklines=true,
        columns=fullflexible
    }
]
\label{app:prompt_template_geography}
\begin{verbatim}

You are playing an epidemic decision game where your goal is to maximise your 
total points.

Game mechanics:
- A disease spreads through a contact network: interacting with others exposes 
you to infection risk.
- Your infection risk increases with local prevalence and your contacts.
- If infected, you may lose points (reduced rewards, possible large penalties).
- You move through health states (susceptible -> infected -> recovered).

Decision each round:
- Quarantine: {low_reward} pts. No infection risk this round.
- Stay active: {high_reward} pts. Risk infection from contacts.

This is a trade-off between:
- Short-term reward (staying active)
- Long-term risk (infection causing point losses)

Your objective:
Maximise your total points over time. Consider expected future losses from 
infection, not just immediate reward.

Your initial beliefs (from pregame survey, scale 1-6, where 1 = weakest, 
6 = strongest):
- Perceived infection risk: how likely you think it is that you will get 
infected.
- Perceived health severity: how serious you think infection would be for 
your health.
- Quarantine self-efficacy: how confident you are that you can successfully 
follow quarantine.
- Quarantine response efficacy: how effective you think quarantine is at 
preventing spread.

Local prevalence (0-1): fraction of contacts infected in the previous timestep.

Geographic context:
- You are located in [Location].
- Your location affects your baseline exposure risk when staying active.
- In places with higher crowding or mobility, even moderate prevalence
can result in infection.
- In less dense areas, exposure risk may be lower, but it is not zero.
- Use this information directly when deciding whether to quarantine.

Your current state:
- Time: {t}
- Status: {status}
- Infection history: {has_been_infected}
- Points: {points}
- Local prevalence: {local_prev}
- Perceived infection risk (1-6): {perceived_infection_risk}
- Perceived health severity (1-6): {perceived_health_severity}
- Quarantine self-efficacy (1-6): {quarantine_self_efficacy}
- Quarantine response efficacy (1-6): {quarantine_response_efficacy}

Use this framework to guide your decision. 
Should you quarantine this round? Reply with only 'yes' or 'no'.
\end{verbatim}
\end{tcolorbox}
\twocolumn
\section{Behavioural Parameterisation}\label{app:behavioural_params}
To align with the human trial, we parametrised agents' decision-making using four core belief factors drawn from the Health Belief Model \citep{champion2008health}:
\begin{itemize}
\item Perceived susceptibility: operationalised as the agent's assigned infection risk.
\item Perceived severity: operationalised as the agent's assessed health severity.
\item Perceived benefits: operationalised as quarantine response efficacy.
\item Perceived barriers: operationalised through self-efficacy scores and experimental group assignment, the latter serving as a proxy for economic cost.
\end{itemize}
To construct a behavioural profile for each agent, we leveraged responses from both the pre-game (S3) and in-game (S4) surveys of the original study, mapping each relevant question (see Appendix~\ref{app:survey-questions}) to one of the four health belief dimensions. Questions scores were averaged to produce a single representative value. Specifically, questions 35 and 41 were averaged to yield perceived susceptibility, questions 36 and 42 to yield perceived severity, questions 37 and 43 to yield self-efficacy, and questions 38 and 44 to yield quarantine response efficacy. Each answer was mapped from a six-point Likert-scale response (a–f) to an ordinal score of 1–6, and responses from both surveys were pooled before averaging, so that each agent's belief state reflected all available self-report data.
For agents with missing responses on a given dimension, the population median for that dimension was substituted; any remaining gaps were filled with a neutral default of 3.0. These four continuous scores were then assigned directly to the corresponding agent attributes at prompt initialisation.
\section{Survey Questions Used for Agent Belief Initialisation}
\label{app:survey-questions}

The following questions were drawn from the pre-game survey (Survey~3) and the
in-game survey (Survey~4). Responses were recorded on a six-point scale
(\emph{a}~=~1 through \emph{f}~=~6). Each question is labelled with its
database identifier (Q\#) for cross-reference with the main text.

\subsection{Pre-Game Survey (Survey 3)}

\begin{enumerate}
    \item[\textbf{Q35.}] How likely do you think it is that you will get infected
    with a respiratory virus (e.g., flu, COVID) in the next three months?\\
    \textit{1 = Virtually impossible \quad 6 = Almost certain}

    \item[\textbf{Q36.}] If you were to get infected with a respiratory virus
    (e.g., flu, COVID) in the next three months, how serious do you think the
    impact would be on your health?\\
    \textit{1 = No impact at all \quad 6 = Very significant impact}

    \item[\textbf{Q37.}] How confident are you that you would be able to
    successfully follow home quarantine if there is a new infectious disease with
    potentially serious health effects (such as COVID-19 at the beginning of the
    pandemic)?\\
    \textit{1 = Not confident at all \quad 6 = Completely confident}

    \item[\textbf{Q38.}] How effective do you think following home quarantine
    would be in preventing the spread of infectious diseases?\\
    \textit{1 = Not effective at all \quad 6 = Extremely effective}
\end{enumerate}

\subsection{In-Game Survey (Survey 4)}

\begin{enumerate}
    \item[\textbf{Q41.}] How likely do you think it is that your avatar will get
    infected with the virtual pathogen while playing with the Epigames app?\\
    \textit{1 = Virtually impossible \quad 6 = Almost certain}

    \item[\textbf{Q42.}] If your avatar gets infected with the virtual pathogen,
    how much do you think this would affect your chances of winning a gift card
    with the point-based lottery at the end of the game?\\
    \textit{1 = Not affected at all \quad 6 = Extremely affected}

    \item[\textbf{Q43.}] How confident are you that you can successfully follow
    the in-game quarantine?\\
    \textit{1 = Not confident at all \quad 6 = Completely confident}

    \item[\textbf{Q44.}] How effective do you think in-game quarantine is in
    preventing the spread of the virtual pathogen?\\
    \textit{1 = Not effective at all \quad 6 = Extremely effective}
\end{enumerate}

\noindent Questions 39 and 40 (gender and school affiliation) were not used for
belief initialisation and are omitted here.

\section{Decision Consistency Analysis}
\label{app:decision_consistency}
An important question we investigated was whether agents always took the same action when presented with identical network state and context. As mentioned in the main text, this is vital to explore if agents behave inconsistently under identical conditions. However, inconsistency may in fact be desirable. Real human decision-making is inherently stochastic, and agents that always produce identical responses under identical conditions may be overly deterministic relative to the population-level behaviour they intend to model. To explore this, we ran $10$ simulations on a small random network of $50$ agents across 6 decision days and recorded whether each agent took the same action on each day when given identical information. For each agent on each day, this yields a set of $n$ binary responses (yes/no) $\mathcal{A} = \{a_1, \dots, a_n\}$. To quantify how often an agent agreed with itself across runs, we define the self-agreement score:

\begin{equation}
    \mathcal{C} = \frac{\sum_{(i,j) \in \mathcal{A} \times \mathcal{A}} \mathbf{1}[a_i = a_j]}{|\mathcal{A}|^2}
\end{equation}

where $\mathcal{A} \times \mathcal{A} = \{(i,j) \mid i,j \in \mathcal{A}\}$ is the Cartesian 
product and $\mathbb{1}[\cdot]$ is the indicator function. Since outcomes are binary and self-pairs are included, 
$\mathcal{C} \in [0.5, 1]$, where $\mathcal{C} = 0.5$ corresponds to a random split of responses and $\mathcal{C} = 1$ indicates perfect consistency.

\end{appendices}


\end{document}